\documentclass{article}
\usepackage{graphicx}
\begin{document}
\begin{center}
{\Large \bf Suppression of the bottleneck in semiconductor
microcavities}\\{\bigskip T.D. Doan and D.B. Tran Thoai
\footnote{Correspondent author}}\\ {\bigskip \it Ho Chi Minh City
Institute of Physics,\\ Vietnam Centre for Natural Science and
Technology,\\ 1 Mac Dinh Chi, Ho Chi Minh City, Viet Nam}\\
\end{center}

\begin{abstract}
The relaxation kinetics of cavity polaritons by scattering with
thermal acoustic phonons is studied within the rate equation
approximation. Numerical results show that a suppression of the
bottleneck of lower polariton states occurs at high polariton
densities. We have found that the long decay time of the
photon-like polaritons, the thin width of the embedded quantum
wells and the small value of exciton-cavity detuning are favorable
for the suppression of the bottleneck.
\end{abstract}
{\small \noindent {\it Keywords:} A. Semiconductors; D.
Electron-phonon interaction; D. Optical properties}

\section{Introduction}

Semiconductor microcavities (MC) with embedded quantum wells have
found a growing interest in recent years. The eigenstates of the
systems as a result of the strong coupling between the MC photon
and quantum well exciton at the same in-plane wave vector are
called cavity polaritons. The Rabi splitting and the dispersion of
these cavity polaritons are readily observable using reflection,
absorption, photoluminescence measurements [1-3]. Recently, many
new phenomena have been intensely studied in MC: super-linear
behaviour in emission intensity [4-6], stimulated polariton
amplifier [7-8] and suppression of the relaxation bottleneck
\cite{Tartakovski2}. To the best of our knowledge, Tartakovski et
al. \cite{Tartakovski2} were the only group, which has clearly
observed this suppression of the relaxation bottleneck of lower
polariton states at higher powers.

In this work, we study the relaxation kinetics of cavity
polaritons by scattering with thermal acoustic phonons. Here we
will present numerical solutions of the rate equations and show
that a suppression of the bottleneck occurs at sufficiently high
concentration of the polaritons.\\

\section{ Relaxation kinetics of cavity polaritons}

Within the two-coupled band model and considering only the 1s
state of the heavy-hole quantum well exciton, the energy for the
upper and lower cavity polariton branches is given by the usual
solutions of the polariton dispersion equations as given in Ref.
\cite{Tassone}:
\begin{equation}
(E^x_k-E^i_k)(E^c_k-E^i_k)=\frac{\hbar^2\Omega^2}{4}
\label{energy1}
\end{equation}
where
\begin{eqnarray}
 E^c_k&=&\sqrt{(E^c_0)^2+\frac{\hbar^2 c^2}{\epsilon_{cav}}k^2}\label{energy2}\\
 E^x_k&=&E^x_0+\frac{\hbar^2k^2}{2\left(m_e+m_h\right)}\label{energy3}
\end{eqnarray}
Here $E^x_0$ is the fundamental exciton energy, $m_e\ (m_h)$ the
electron (hole) in-plane mass, $n_{cav}$ the cavity refraction
index, $\Omega$ the Rabi splitting. Fig. 1 shows the cavity
polariton dispersion of GaAs quantum wells embedded in cavity for
a detuning $\delta=E^c_0-E^x_0=-2$ meV, a Rabi splitting
$\hbar\Omega=4.8$ meV, and $n_{cav}=\sqrt{\epsilon_{cav}}=3.07$.

In the following we will use the semi-classical Boltzmann kinetics
to investigate the cavity polariton relaxation by acoustic phonon
scattering. The rate equation for the distribution $f^i_k$ of
branch i with the in-plane wave vector k and energy $E^i_k$ of the
cavity polariton can be written as [10-11]
\begin{eqnarray}
  \frac{\partial}{\partial t}f^i_k =
  -\frac{f^i_k}{\tau^i_k}-\sum_{j,\overrightarrow{k}}\left[W^{i,j}_{\overrightarrow{k},\overrightarrow{k}'}
  f^i_k \left(1+f^j_{k'}\right)-\left(\overrightarrow{k},i\rightleftharpoons\overrightarrow{k}',j\right)\right]+G^i_k(t)
  \label{reqn}
\end{eqnarray}
$\frac{1}{\tau^i_k}$ is the radiative recombination rate and
$G^i_k (t)\equiv A^i_k H(t)$ the generation rate. The transition
rates $W^{i,j}_{\overrightarrow{k},\overrightarrow{k}'}$ due to
deformation potential scattering are weighted with exciton
Hopfield coefficients $x^i_k$ and $x^j_{k'}$:
\begin{eqnarray}
  W^{i,j}_{\overrightarrow{k},\overrightarrow{k}'}&=&\frac{L_z\left(x^i_kx^j_{k'}\right)^2}{\hbar\rho V u^2q_z}
  \left(\Delta^{i,j}_{k,k'}\right)^2A^2(q_z)D^2(|\overrightarrow{k}-\overrightarrow{k}'|)
  \nonumber\\
  &&\left|N^p_{E^j_{k'}-E^i_k}\right|\Theta\left(\Delta^{i,j}_{k,k'}-|\overrightarrow{k}-\overrightarrow{k}'|\right)
  \label{trans.rate}
\end{eqnarray}
where
\begin{eqnarray}
  &&\Delta^{i,j}_{k,k'}=\frac{\left|E^i_k-E^j_{k'}\right|}{\hbar u}\label{del}\\
  &&q_z=\sqrt{\left(\Delta^{i,j}_{k.k'}\right)^2-|\overrightarrow{k}-\overrightarrow{k}'|^2}\label{qz}\\
  &&D(x)=D_eG\left(\frac{xm_h}{m_e+m_h}\right)-D_hG\left(\frac{xm_e}{m_e+m_h}\right)\label{f}\\
  &&A(x)=\frac{8\pi^2}{L_z x\left(4\pi^2-L_z^2 x^2\right)}sin\left(\frac{L_z x}{2}\right)\label{f2}\\
  &&G(x)=\frac{1}{\left[1+\left(x a_b/2\right)^2\right]^{3/2}}\  .\label{f1}
\end{eqnarray}
Here $L_z$ is quantum well width, V the sample volume, u the
longitudinal sound velocity , $\rho$ the mass density of the
solid, $D_e$ and $D_h$ the deformation potentials. A(x) and G(x)
are the Fourier transform of $\chi^2_e(z_e)$ and $F^2(\rho)$,
respectively. Note that the 1s exciton wave function has been
taken in the simplest form
\begin{equation}
  \phi^{1s}_{x}(r_e,r_h)=F(\rho)\chi_e(z_e)\chi_h(z_h)\frac{1}{\sqrt{S}}e^{i\overrightarrow{k}\overrightarrow{R}}.
  \label{exc.func}
\end{equation}
Expression (\ref{trans.rate}) contains transition rate due to the
absorption of phonon $\propto\ N^p_E$ for $E > 0$, and due to
emission of phonon $\propto\ \left(N^p_{|E|}+1\right)$ for $E <
0$, where $N^p_{E > 0} = \left(e^{\beta E}-1\right)^{-1}$ is the
thermal phonon distribution function.

In the following we will present the results of the cavity
polariton kinetics on the lower branch assuming an isotropic
distribution of polaritons in k space. For the decay rate
$\frac{1}{\tau^i_k}$, we use the same approximation given by Bloch
and Marzin in GaAs quantum well \cite{Bloch}:
$\frac{1}{\tau^i_k}=\frac{\left(c^i_k\right)^2}{\tau^c}$ for $0 <
k < k_{cav}=6 \times 10^4\ cm^{-1}$;
$\frac{1}{\tau^i_k}=\frac{1}{\tau^x}$ for
$k_{cav}<k<k_{rad}=\frac{n_{cav}E^x_0}{\hbar c}=2.1\times 10^5\
cm^{-1}$; and $\frac{1}{\tau^i_k}=0$ for $k>k_{rad}$, ($\tau^c$
and $\tau^x$ are the radiative life of photon and exciton,
respectively, $c^i_k$ the photon Hopfield coefficient).\\

\section{Numerical results for kinetics of cavity polaritons}

In this section, rate equations are numerically integrated using
the following parameters for MC with embedded GaAs quantum wells:
[12-13] $E^x_0=1.3955\ eV,\ m_e=0.067 m_0,\ m_h=0.13 m_0,\
\hbar\Omega=4.8\ meV,\ a_b=7.5\ nm,\ D_h=12\ eV,\ D_e=-7\ eV$
\cite{wolfe}, $u=4.81\times 10^3\ m/s,\ \rho=5.3\times 10^3\
kg/m^3,\ n_{cav}=3.07,\ \tau_x=200\ ps$. Concerning the generation
rate we choose for $A^i_k$ a Gausian distribution of exciton on
the lower branch with a width $\Delta E=0.1\ meV$ centered at a
momentum $k_0=5\times 10^5\ cm^{-1}$ and $H(t)=H_0\
th\left(\frac{t}{T}\right)$ with $T=20\ ps$. Unless otherwise
stated we will choose $L_z=6\ nm,\ T_{phon}=2\ K$,
$\delta=E^c_0-E^x_0=-1\ meV$.

In Fig. 2 we show the distribution function $f_k(t)$ of cavity
polaritons on the lower branch for a rather long radiative photon
decay time $\tau_c=50\ ps$ at $n=7.8\times 10^9\ cm^{-2}$. We
clearly see the presence of the bottleneck effect similar to the
bulk case [15-16]. However, if we increase the pump intensity, a
suppression of the bottleneck occurs as can be clearly seen in
Fig. 3 at $n=5.1\times 10^{10}\ cm^{-2}$. These findings agree
rather well with recent measurements performed in GaAs cavity by
Tartakovski et al. \cite{Tartakovski2} using nonresonant cw laser
excitation. They found that the bottleneck has been strongly
suppressed at higher power laser excitation. This feature is
specific for cavity polaritons. We have also carried out the
calculations for the ortho-exciton polaritons in bulk $Cu_2O$
\cite{Bao} and have found that the bottleneck persists up to very
high density around $10^{18}\ cm^{-3}$.The main difference in the
dispersion relation between cavity polaritons and bulk polaritons
is the slope of $E_k$ versus k in the bottleneck k-region. The
very high value of the slope of the bulk polaritons (70 times
higher than the slope of the cavity polaritons) is very
unfavorable for the suppression of the bottleneck even at very
high densities. On the other hand, one can lower the slope in the
cavity polaritons by decreasing the absolute value of the negative
detuning $\delta$. From expression (\ref{f2}) of $A(x)$ we can see
that the scattering rate is roughly proportional to
$\left(sin\frac{L_z}{2}/L_z\right)^2$. Therefore, quantum wells
with thin width will be favorable for the suppression of the
bottleneck. For $L_z=3\ nm,\ \delta=-0.25\ meV $ and $\tau_c=50\
ps$ the bottleneck starts to be suppressed at densities, which are
roughly 3 time lower than those for $L_z=6\ nm,\ \delta=-1\ meV $
and $\tau_c=50\ ps$ . For a more realistic $\tau_c$ one needs a
higher density to compensate the fast decay of the photon-like
polaritons. Recently, Senellart and Bloch have given the photon
decay time $\tau_c$ a value of $6.6\ ps$ \cite{Senellart}. In Fig.
4 we display the integrated distribution function
$\left(\int_0^{\infty}dt f_k(t)\right)$ versus the wave number k
for different polariton densities for $\tau_c=6.6\ ps$, $L_z=3\
nm,\ \delta=-0.25\ meV$ at 2K (Fig. 4a) and 15K (Fig. 4b). These
results show a clear process of the relaxation kinetics: a
nonlinear increase of low k cavity polaritons with increasing
polariton densities (or excitation laser powers) leads to the
suppression of the bottleneck. The suppression is slightly more
favorable at T=15 K. For density $n> 10^{10}\ cm^{-2}$, one should
take into account the exciton-exciton scattering and also go
beyond the rate equation approximation. However, we expect our
results remain qualitatively valid even in a better treatment of
polariton relaxation kinetics.

In conclusion, we have presented detailed cavity polariton
relaxation kinetics of lower branch due to deformation potential
scattering by accoustic phonons. In agreement with recent
experimental results by Tartakovski et al. \cite{Tartakovski2} we
found that suppression of the bottleneck is found at high
polariton densities.\\

\noindent{\bf{Acknowledgements}} \vspace{.2cm}

We gratefully acknowledge the financial support of the National
Program for Basic Research.

\newpage
\thispagestyle{empty}
\noindent{\bf Figure captions:}\\ \\
 {\bf Fig. 1:} Dispersion of cavity polaritons with embedded GaAs quantum wells for
 $\delta=-2\ meV$, $\hbar\Omega=4.8\ meV$. \\ \\
 {\bf Fig. 2:} Resulting distributions of cavity polaritons on lower branch versus time for
 $\delta=-1\ meV,\ L_z=6\ nm,\ \tau_c=50\ ps$, $n=7.8\times 10^{9}\ cm^{-2}$ at a bath temperature of 2 K.  \\ \\
 {\bf Fig. 3:} Suppression of the bottleneck at $n=5.1\times 10^{10}\ cm^{-2}$ for $\delta=-1\ meV,\ L_z=6\ nm,\ \tau_c=50\ ps$
 and $T=2\ K$. The inset shows total polariton density.\\ \\
 {\bf Fig. 4:} Resulting integrated distribution function versus wave number k
 for different polariton densities at 2K (a) and 15K (b) for $\delta=-0.25\ meV,\ L_z=3\ nm,\ \tau_c=6.6\ ps$. \\ \\

\newpage
\thispagestyle{empty}
\includegraphics[width=11cm,angle=-90]{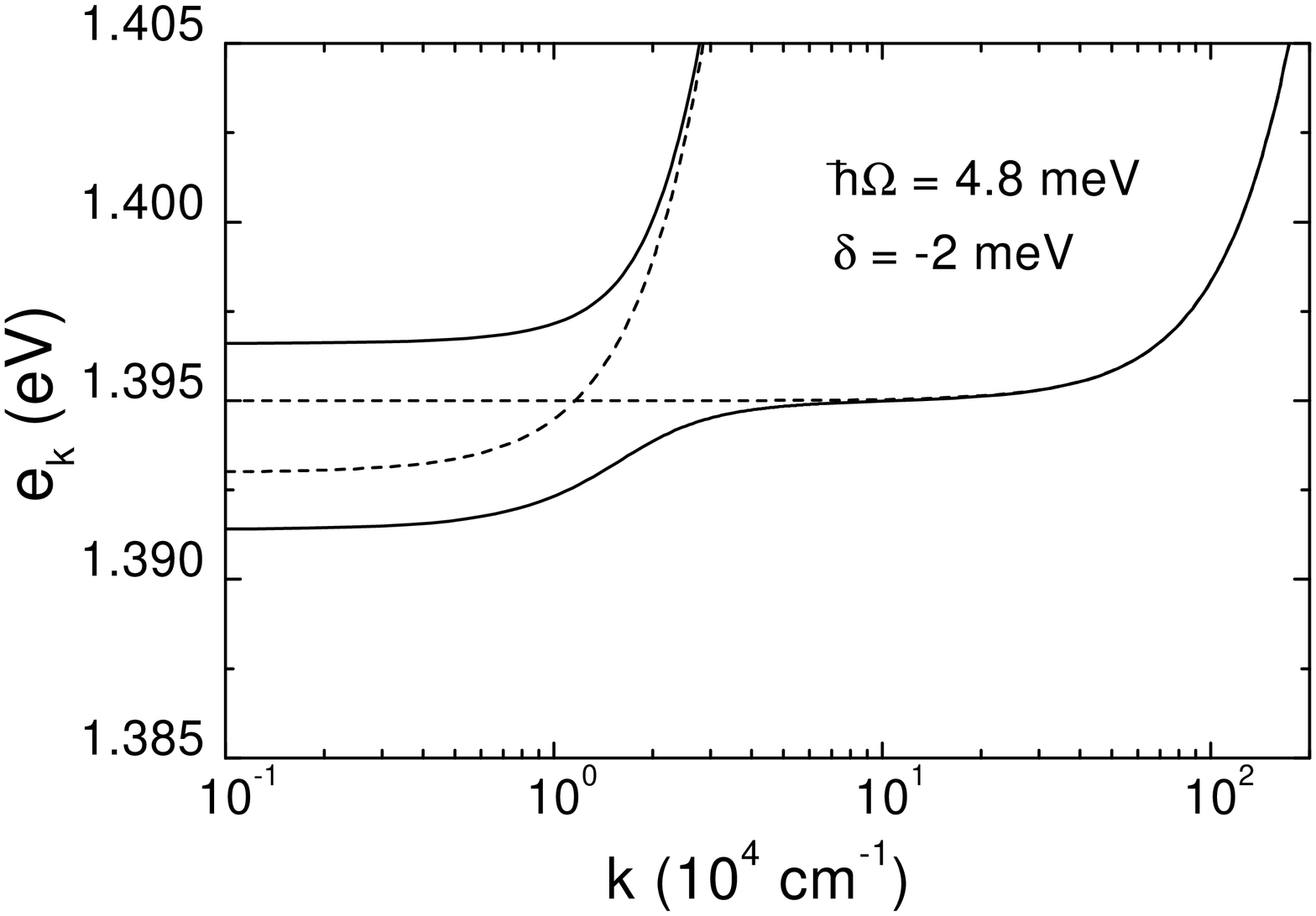}\\
{\large{\bf Fig. 1}: Dispersion of cavity polaritons with embedded
GaAs quantum wells for $\delta=-2\ meV$, $\hbar\Omega=4.8\ meV$.}

\newpage
\thispagestyle{empty}
\includegraphics[width=14.5cm]{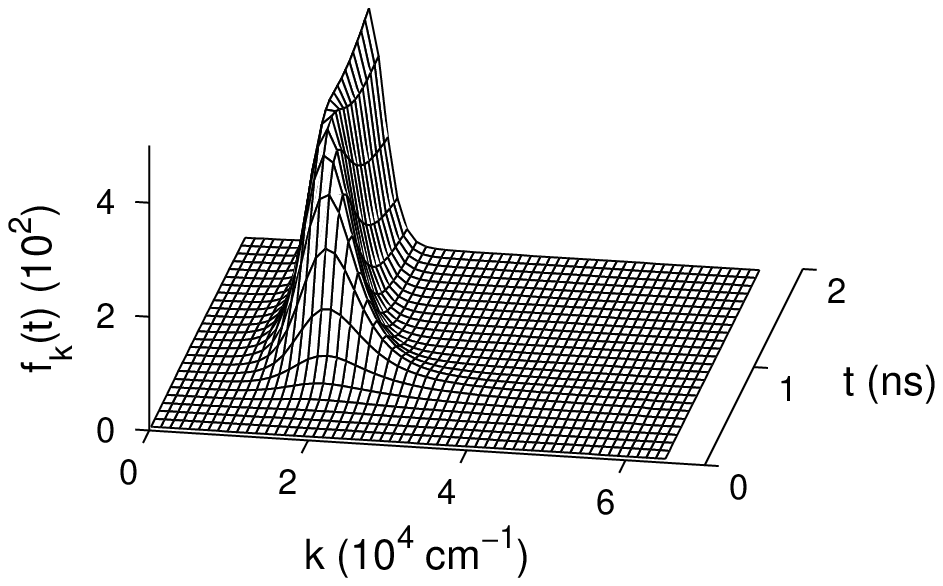}\\ \\
{\large{\bf Fig. 2}: Resulting distributions of cavity polaritons
on lower branch versus time for $\delta=-1\ meV,\ L_z=6\ nm,\
\tau_c=50\ ps$, $n=7.8\times 10^{9}\ cm^{-2}$ at a bath
temperature of 2 K.}

\newpage
\thispagestyle{empty}
\includegraphics[width=14cm]{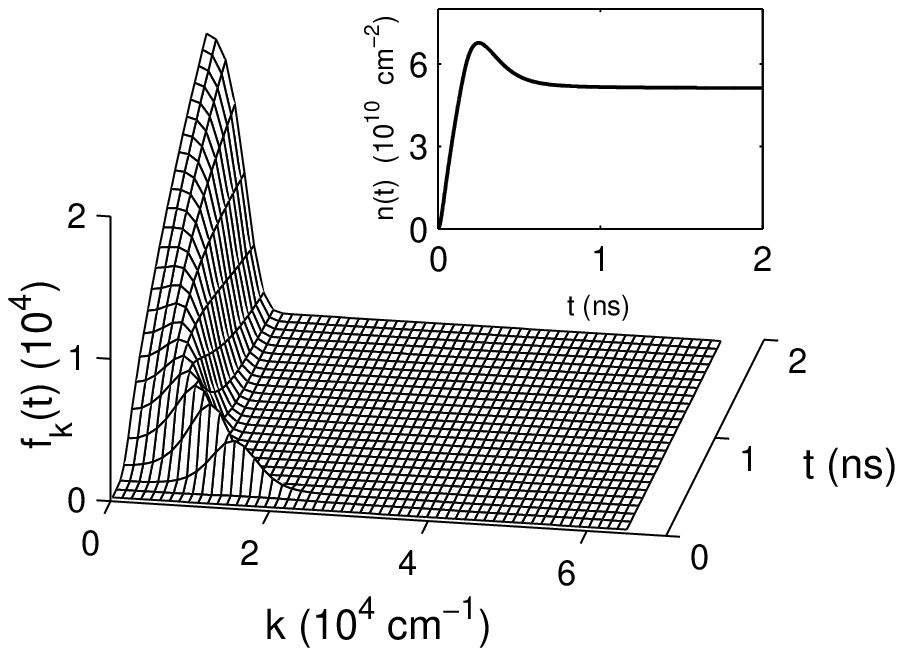}\\ \\
{\large{\bf Fig. 3}: Suppression of the bottleneck at $n=5.1\times
10^{10}\ cm^{-2}$ for $\delta=-1\ meV,\ L_z=6\ nm,\ \tau_c=50\ ps$
and $T=2\ K$. The inset shows total polariton density.}

\newpage
\thispagestyle{empty}
\includegraphics[width=14 cm]{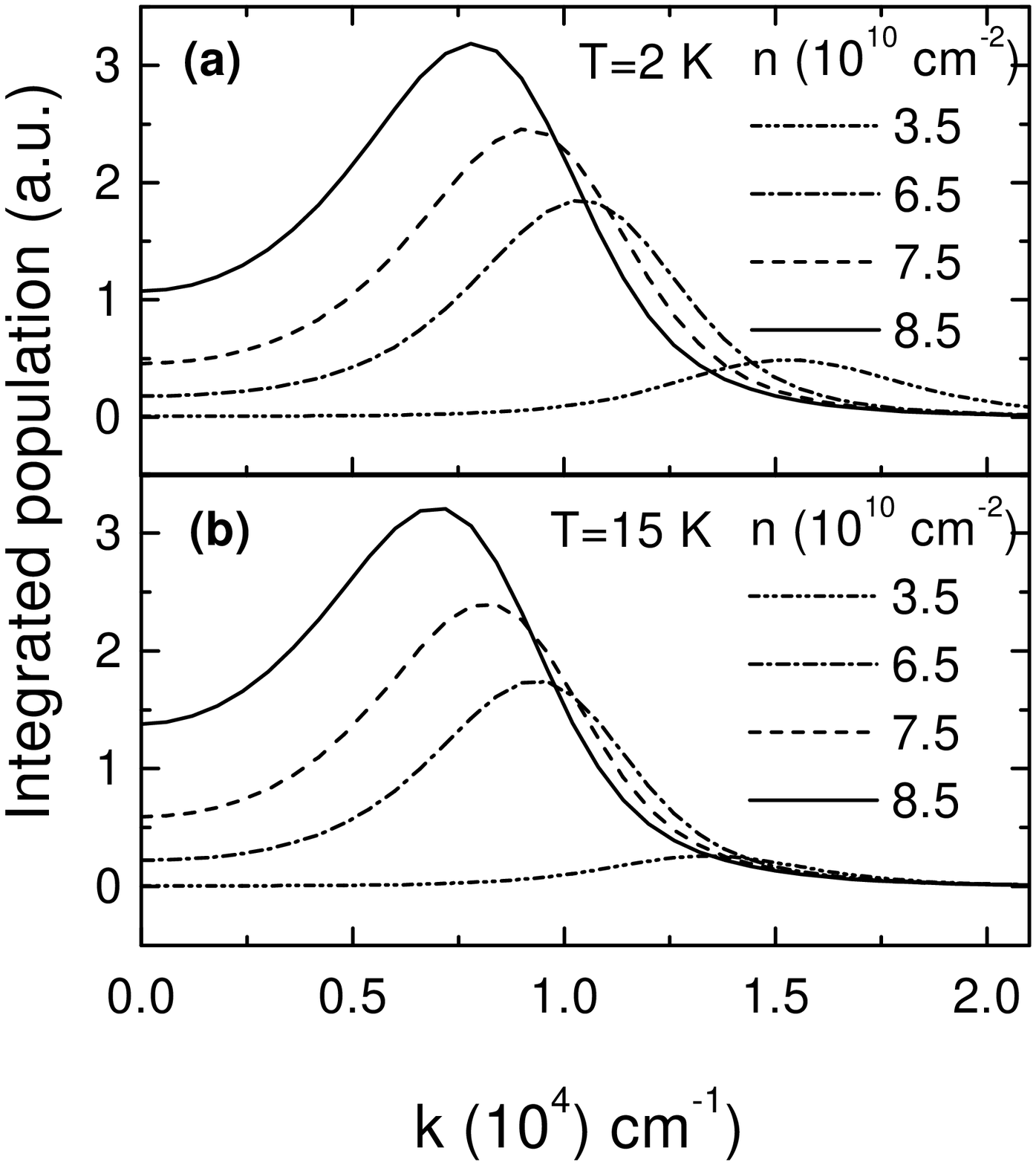}
{\large{\bf Fig. 4}: Resulting integrated distribution function
versus wave number k for different polariton densities at 2K (a)
and 15K (b) for $\delta=-0.25\ meV,\ L_z=3\ nm,\ \tau_c=6.6\ ps$.}

\end{document}